\documentclass[twocolumn,showpacs,prb,preprintnumbers,superscriptaddress,amsmath,amssymb]{revtex4}
\usepackage{graphicx}% Include figure files
\usepackage{dcolumn}% Align table columns on decimal point
\usepackage{bm}% bold math
\usepackage{amsmath,amssymb}
\newcommand{\be}{\begin{eqnarray}}
\newcommand{\ee}{\end{eqnarray}}

\begin{document}
\title{Exact solution to an interacting dimerized Kitaev model at symmetric point}
\author{Yucheng Wang}
\affiliation{Beijing National
Laboratory for Condensed Matter Physics, Institute of Physics,
Chinese Academy of Sciences, Beijing 100190, China}
\affiliation{School of Physical Sciences, University of Chinese Academy of Sciences, Beijing, 100049, China}
\author{Jian-Jian Miao}
\affiliation{Kavli Institute for Theoretical Sciences,
Chinese Academy of Sciences, University of Chinese Academy of Sciences, Beijing, 100190, China}
\author{Hui-Ke Jin}
\affiliation{Department of Physics, Zhejiang University, Hangzhou 310027, China}
\affiliation{Collaborative Innovation Center of Advanced Microstructures, Nanjing 210093, China}
\author{Shu Chen}
\thanks{schen@iphy.ac.cn}
\affiliation{Beijing National
Laboratory for Condensed Matter Physics, Institute of Physics,
Chinese Academy of Sciences, Beijing 100190, China}
\affiliation{School of Physical Sciences, University of Chinese Academy of Sciences, Beijing, 100049, China}
\affiliation{Collaborative Innovation Center of Quantum Matter, Beijing, China}
\begin{abstract}
%A dimerized Kitaev model is comprised of the Su-Schrieffer-Heeger (SSH) model and Kitaev model
We study the interacting dimerized Kitaev chain at the symmetry point $\Delta=t$ and the chemical potential $\mu=0$ under open boundary conditions, which can be exactly solved by applying two Jordan-wigner transformations and a spin rotation. By using exact analytic methods, we calculate two edge correlation functions of Majorana fermions and demonstrate that they can be used to distinguish different topological phases and characterize the topological phase transitions of the interacting system. According to the thermodynamic limit values of these two edge correlation functions, we give the phase diagram of the interacting system which includes three different topological phases: the trivial, the topological superconductor and the Su-Schrieffer-Heeger-like topological phase and we further distinguish the trivial phase by obtaining the local density distribution numerically.
\end{abstract}
\pacs{74.20.-z, 74.78.-w, 05.30.Rt, 71.10.Pm}
%74.20.-z:theories and models of superconducting state; 74.78.-w:superconduting films and low-dimensional structures;
%05.30.Rt: quantum phase transitions; 71.10.Pm: fermions in reduced dimensions .
\maketitle
%%%%%%%%%%%%%%%%%%%%%%%%
\section{Introduction}
%%%%%%%%%%%%%%%%%%%%%%%%%
Majorana zero modes have attracted intensive studies in past years \cite{Alicea,Gangadharaiah,Stoudenmire,Chan,Sela}. This is not only due to their potential applications in topological quantum computation \cite{Nayak}, but also some reported experimental evidences of their existence \cite{Kane,Alicea2,Sarma2,Fujimoto,Sarma3,Mourik,Yazdani,Deng}. Kitaev chain model \cite{Kitaev} provides a simple platform to study the Majorana zero modes, which has recently attracted a lot of attentions \cite{Wilczek,Elliott}. The topological phase transitions of this model can be obtained by calculating the Majorana number \cite{Kitaev,Lang1} under periodic boundary conditions or calculating an edge correlation function \cite{Miao1} under open boundary conditions (OBC). On the other hand, as the simplest example of 1D topological insulators, the Su-Schrieffer-Heeger (SSH) model \cite{ssh}, or known as a dimerized one-dimensional (1D) model, has been an important model system to show rich topological phenomena \cite{ssh2,ssh3,Maki,Sarma,Linhu}. A dimerized Kitaev model \cite{Nagaosa,Chen} comprised of the SSH model and the Kitaev model can give rise to a rich phase diagram, which includes the trivial phase, topological superconductor (TSC) phase and SSH-like topological phase.

The effect of interactions on topological insulators \cite{Zhang} or superconductors remains an open problem. The Kitaev model with interaction in some special cases has been widely studied analytically \cite{Kitaev2,Turner,Goldstein,Hassler} and numerically \cite{Thomale,Rahmani,Milsted,Iemini,Gergs}. Recent work by Miao et al. \cite{Miao2} found the exact solution in the symmetric region ($\mu=0, \Delta=t$) of the Kitaev chain with nearest neighbor interaction and showed the phase transition between TSC phase and trivial phase by introducing an edge correlation function of Majorana fermions. McGinley et al. \cite{McGinley} further extended the exact solution to the disorder case in this symmetric region. In this work, we study the dimerized Kitaev model with nearest neighbor interaction at the symmetric point of $\mu=0$ and $\Delta=t$ under OBC with the help of exact solution by mapping the interacting model onto an noninteracting fermion models after two Jordan-wigner transformations and a spin rotation. One of our motivations is to see the effect of interaction on the trivial, TSC and SSH-like topological phases in the noninteracting dimerized Kitaev model, and give an exact phase diagram of the interacting dimerized Kitaev model at the symmetric point. In order to distinguish different phases, we introduce two edge correlation functions and use them to distinguish different topological phases of the interacting dimerized Kitaev model. We find that the trivial phase, TSC phase and SSH-like topological phase of this interacting system can be well distinguished by these two edge correlation functions and the phase transition points can be analytically obtained.

This paper is organized as follows: in Sec. \ref{model}, we introduce the interacting dimerized Kitaev superconductor model and it's form in the Majorana representation. In Sec. \ref{diagonalization}, the Hamiltonian is diagonalized by using two Jordan-Wigner transformations and the singular value decomposition (SVD). In Sec. \ref{edge} we introduce two edge correlation functions and obtain the phase diagram, and then we further distinguish the trivial phase by using the energy spectra and the local density distribution. A brief summary is given in Sec. \ref{summary}.

%%%%%%%%%%%%%%%%%%%%%%%%%%%%%%%%%%%%%%%
\section{Model Hamiltonian}
\label{model}
%%%%%%%%%%%%%%%%%%%%%%%%%%%%%%%%%%%%%%
We consider an interacting dimerized Kitaev superconductor chain under OBC, which is described by
\begin{widetext}
\begin{eqnarray}
H&=&\sum_{j} \left\{-t[(1+\eta)c_{j,A}^{\dagger}c_{j,B}+(1-\eta)c_{j,B}^{\dagger}c_{j+1,A}+h.c. ] - \Delta [(1+\eta)c_{j,A}^{\dagger}c_{j,B}^{\dagger}+(1-\eta)c_{j,B}^{\dagger}c_{j+1,A}^{\dagger}+h.c. ] \right. \nonumber\\
&& \left. +  4U \left[ (n_{j,A}-\frac{1}{2})(n_{j,B}-\frac{1}{2}) + (n_{j,B}-\frac{1}{2})(n_{j+1,A}-\frac{1}{2}) \right] \right \}
 -\mu\sum_{j} (c_{j,A}^{\dagger}c_{j,A}+c_{j,B}^{\dagger}c_{j,B}) ,
\label{ham-1}
\end{eqnarray}
\end{widetext}
where $c_{j,A}, c_{j,B} (c_{j,A}^{\dagger}, c_{j,B}^{\dagger})$ are fermion annihilation (creation) operators on site $A, B$ of the $jth$ cell respectively, $n_{j,A}=c_{j,A}^{\dagger}c_{j,A}$ and $n_{j,B}=c_{j,B}^{\dagger}c_{j,B}$ are the corresponding fermion occupation number operators,
$t$ denotes the hopping integral, $\Delta$ is the superconducting pairing gap taken to be real, $\mu$ is the chemical potential, and $U$ denotes the nearest neighbor interaction. We set $|\eta|<1$ and the number of cell $L_c$, which equals $L_s/2$, where $L_s$ is the length of this chain. When $U=0$, $\Delta=0$ and $\mu=0$, the Hamiltonian is reduced to the SSH model \cite{ssh}; when $U=0$ and $\eta=0$, the Hamiltonian is reduced to the 1D kitaev model \cite{Kitaev}, and when $U=0$, this model is a dimerized Kitaev model \cite{Nagaosa,Chen}.

The particle-hole conjugation operator $Z_{2}^{p}$ defined as $Z_{2}^{p}=\prod_{j}\left[c_{j}+\left(-1\right)^{j}c_{j}^{\dagger}\right]$ \cite{Miao2} and one can easily verify that $(Z_{2}^{p})^{-1}c_jZ_{2}^{p}=(-1)^{j}c_{j}^{\dagger}$. $Z_{2}^p$ is conserved when $\mu=0$, i.e., this system has the particle-hole symmetry when $\mu=0$. Next, we shall study the interacting dimerized Kitaev model at the symmetric point of $\mu=0$ and $\Delta=t$.
%Another good quantum number is the fermion number parity $Z_{2}^{f}$ defined as,
%\begin{equation}\label{def:Z2f}
%Z_{2}^{f}=e^{i\pi\sum_{j}n_{j}}=\left(-1\right)^{\hat{N}},
%\end{equation}
%where $\hat{N}=\sum_{j}n_{j}$ is the number of fermions in the system.  It is obvious that $(Z_{2}^f)^2=1$ and $[H,Z_{2}^f]=0$.
%%Note that $Z_{2}^{f}$ conserves in the whole parameter space.%

We introduce the Majorana fermion operators $c_{j,\delta} = \frac{1}{2}\left(\gamma_{j,\delta}^{a}+i\gamma_{j,\delta}^{b}\right)$ and
$c_{j,\delta}^{\dagger} = \frac{1}{2}\left(\gamma_{j,\delta}^{a}-i\gamma_{j,\delta}^{b}\right)$, where $\delta=A, B$. The Majorana fermion operators should be real
$\left(\gamma_{j,\delta}^{\beta}\right)^{\dagger}=\gamma_{j,\delta}^{\beta}$, where $\beta=a, b$, and they fulfill the anticommutation relations
$\left\{\gamma_{j}^{\beta},\gamma_{l}^{\beta'}\right\} =2\delta_{\beta\beta'}\delta_{jl}$,
where $\beta'=a, b$. By using the Majorana operators, the Hamiltonian
of the interacting dimerized Kitaev chain becomes
\begin{widetext}
\begin{eqnarray}
&H=\frac{i}{2}&\sum_{j} [-(t+\Delta)(1+\eta)\gamma_{j,B}^{a}\gamma_{j,A}^{b}-(t-\Delta)(1+\eta)\gamma_{j,A}^{a}\gamma_{j,B}^{b}
-(t+\Delta)(1-\eta)\gamma_{j+1,A}^{a}\gamma_{j,B}^{b}-(t-\Delta)(1-\eta)\gamma_{j,B}^{a}\gamma_{j+1,A}^{b}\nonumber\\
&&-U\gamma_{j,A}^{a}\gamma_{j,A}^{b}\gamma_{j,B}^{a}\gamma_{j,B}^{b}-U\gamma_{j,B}^{a}\gamma_{j,B}^{b}\gamma_{j+1,A}^{a}\gamma_{j+1,A}^{b}].
\label{ham-2}
\end{eqnarray}
\end{widetext}

%%%%%%%%%%%%%%%%%%%%%%%%%%%%%%%%%%%%%%%
\section{Exact diagonalization}
\label{diagonalization}
%%%%%%%%%%%%%%%%%%%%%%%%%%%%%%%%%%%%%%
%%%%%%%%%%%%%%%%%%%%%%%%%%%%%%%%%%%%%%%
\subsection{Mapping to non-interacting chain}
%%%%%%%%%%%%%%%%%%%%%%%%%%%%%%%%%%%%%%
By using two Jordan-Wigner transformations \cite{Jordan,Fisher} and a spin rotation \cite{Miao2}, one can map the Hamiltonian (\ref{ham-2}) to a non-interacting model at the symmetric point of $\Delta=t$ and $\mu=0$.
Firstly, the Hamiltonian (\ref{ham-2}) can be mapped to a typical $XZ$ model by introducing the Jordan-Wigner transformation that $\sigma_{j,\delta}^{x} = \gamma_{j,\delta}^{a}e^{i\pi\sum_{l<j}n_{l}}$,
$\sigma_{j,\delta}^{y}  =  -\gamma_{j,\delta}^{b}e^{i\pi\sum_{l<j}n_{l}}$ and
$\sigma_{j,\delta}^{z}  =  i\gamma_{j,\delta}^{a}\gamma_{j,\delta}^{b}$,
where $\delta=A, B$. Then the Hamiltonian (\ref{ham-2}) can be written as
\begin{eqnarray}
H&=&\sum_{j}[-t(1+\eta)\sigma_{j,A}^{x}\sigma_{j,B}^{x}-t(1-\eta)\sigma_{j,B}^{x}\sigma_{j+1,A}^{x}\nonumber \\
&&+U\sigma_{j,A}^{z}\sigma_{j,B}^{z}+U\sigma_{j,B}^{z}\sigma_{j+1,A}^{z}]
\label{XZ}
\end{eqnarray}

Secondly, we introduce the rotation operator $R=e^{-i\frac{\pi}{4}\sum_{j}\sigma_{j}^{x}}$, which means all the spins are rotated $\frac{\pi}{2}$ around the $x$-axis.
Therefore the $XZ$ chain becomes a $XY$ chain,
\begin{eqnarray}
H&=&\sum_{j}[-t(1+\eta)\tilde{\sigma}_{j,A}^{x}\tilde{\sigma}_{j,B}^{x}-t(1-\eta)\tilde{\sigma}_{j,B}^{x}\tilde{\sigma}_{j+1,A}^{x}\nonumber \\
&&+U\tilde{\sigma}_{j,A}^{y}\tilde{\sigma}_{j,B}^{y}+U\tilde{\sigma}_{j,B}^{y}\tilde{\sigma}_{j+1,A}^{y}].
\end{eqnarray}
where $\tilde{\sigma}_{j}^{x}:= R \sigma_{j}^{x} R^{-1}=\sigma_{j}^{x}$ and $\tilde{\sigma}_{j}^{y}:=R \sigma_{j}^{y} R^{-1}=\sigma_{j}^{z}$.

Finally, we use the Jordan-Wigner transformation again,
$\tilde{\sigma}_{j,\delta}^{x}  =  \tilde{\gamma}_{j,\delta}^{a}e^{i\pi\sum_{l<j}\tilde{n}_{l}}$,
$\tilde{\sigma}_{j,\delta}^{y}  =  -\tilde{\gamma}_{j,\delta}^{b}e^{i\pi\sum_{l<j}\tilde{n}_{l}}$ and
$\tilde{\sigma}_{j,\delta}^{z}  =  i\tilde{\gamma}_{j,\delta}^{a}\tilde{\gamma}_{j,\delta}^{b}$,
to transform the $XY$ chain to a quadratic fermion Hamiltonian \cite{Lieb}, which is written as
\begin{equation}\label{ham-3}
H=\frac{i}{2}\sum_{j,l=1}^{L_s}\tilde{\gamma}_{j}^{a}B_{jl}\tilde{\gamma}_{l}^{b},
\end{equation}
where we set $\tilde{\gamma}_{j,A}^{\beta}=\tilde{\gamma}_{2j-1}^{\beta}$ and $\tilde{\gamma}_{j,B}^{\beta}=\tilde{\gamma}_{2j}^{\beta}$. Here $B_{j,j+1}=2U$, $B_{2j,2j-1}=-2t(1+\eta)$ and $B_{2j+1,2j}=-2t(1-\eta)$.
One can verify that $\tilde{\gamma}_{j}^{a,b}$ are Majorana fermion operators \cite{Miao2}, which satisfy
$\left(\tilde{\gamma}_{j}^{a}\right)^{\dagger}=\tilde{\gamma}_{j}^{a}$, $\left(\tilde{\gamma}_{j}^{b}\right)^{\dagger}=\tilde{\gamma}_{j}^{b}$ and $\left\{\tilde{\gamma}_{j}^{\beta},\tilde{\gamma}_{l}^{\beta'}\right\} =2\delta_{\beta\beta'}\delta_{jl}$. Thus, the Hamiltonian (\ref{ham-2}) is mapped to a non-interacting fermion Hamiltonian (\ref{ham-3})
when $\Delta=t$ and $\mu=0$.

%%%%%%%%%%%%%%%%%%%%%%%%%%%%%%%%%%%%%%%
\subsection{Exact diagonalization}
%%%%%%%%%%%%%%%%%%%%%%%%%%%%%%%%%%%%%%
 The Hamiltonian (\ref{ham-3}) can be exactly diagonalized by using the SVD, i.e.,
the matrix $B$ given in Eq.~\ref{ham-3} can be written as $B=U\Lambda V^T$ \cite{Miao1,Lieb,Katsura}, where $\Lambda$ is a real diagonal matrix whose diagonal elements $\Lambda_k$ are the singular values of $B$.
$U$ and $V$ are two real orthogonal matrices and transform the Majorana operators as
$\tilde{\gamma}_{k}^{a} = \sum_{j=1}^{L_s}U_{jk}\tilde{\gamma}_{j}^{a}$ and $\tilde{\gamma}_{k}^{b} = \sum_{j=1}^{L_s}V_{jk}\tilde{\gamma}_{j}^{b}$.
Similarly, we have
$\left(\tilde{\gamma}_{k}^{\beta}\right)^{\dagger}=\tilde{\gamma}_{k}^{\beta}$ and $\left\{ \tilde{\gamma}_{k}^{\beta},\tilde{\gamma}_{p}^{\beta'}\right\} =2\delta_{\beta\beta'}\delta_{kp}$.

The Hamiltonian can be diagonalized as
\begin{equation}
H=\frac{i}{2}\sum_{k}\tilde{\gamma}_{k}^{a}\Lambda_{k}\tilde{\gamma}_{k}^{b}=\sum_{k}\Lambda_{k}\left(\tilde{c}_{k}^{\dagger}\tilde{c}_{k}-\frac{1}{2}\right),
\end{equation}
where $\tilde{c}_{k}$ and $\tilde{c}_{k}^{\dagger}$ are fermion operators, which fulfill $\tilde{c}_{k}=\frac{1}{2}\left(\tilde{\gamma}_{k}^{a}+i\tilde{\gamma}_{k}^{b}\right)$ and $\tilde{c}_{k}^{\dagger}=\frac{1}{2}\left(\tilde{\gamma}_{k}^{a}-i\tilde{\gamma}_{k}^{b}\right)$. There exist two non-negative singular values for each $k$, which are
\begin{widetext}
\begin{eqnarray}
\Lambda_{k^{I}}=\sqrt{4t^2(1+\eta)^2+4U^2-8tU(1+\eta)\cos(2k^{I})}.\\
\Lambda_{k^{II}}=\sqrt{4t^2(1-\eta)^2+4U^2-8tU(1-\eta)\cos(2k^{II})}.
\end{eqnarray}
\end{widetext}
One can see that $\Lambda_{k^{I}}\geq 0$ and $\Lambda_{k^{II}}\geq 0$, $\Lambda_{k^{I}}= 0$ when $U= t(1+\eta)$ and $k^{I}= 0$ or when $U= -t(1+\eta)$ and $k^{I}= \frac{\pi}{2}$, and $\Lambda_{k^{II}}= 0$ when $U= t(1-\eta)$ and $k^{II}= 0$ or when $U= -t(1-\eta)$ and $k^{II}= \frac{\pi}{2}$. Therefore the spectrum $\Lambda_{k}$ is gap closed at the cases that $U=\pm t(1+\eta)$ and $U=\pm t(1-\eta)$.

For the $\Lambda_{k^{I}}$ and $\Lambda_{k^{II}}$, the corresponding $U$ and $V$ are
\begin{subequations}
\begin{eqnarray}
U_{jk^{I}} & = & \begin{cases}
0, & j = odd,\\
A_{k^{I}}\sin jk^{I}, & j = even,
\end{cases}\\
V_{jk^{I}} & = & \begin{cases}
-A_{k^{I}}\delta_{k^{I}}\sin \left(L_s+1-j\right)k^{I}, & j=odd,\\
0, & j = even.
\end{cases}
\end{eqnarray}
\end{subequations}
and
\begin{subequations}
\begin{eqnarray}
U_{jk^{II}} & = & \begin{cases}
A_{k^{II}}\sin \left(L_s+1-j\right)k^{II}, & j=odd,\\
0, & j=even,
\end{cases}\\
V_{jk^{II}} & = & \begin{cases}
0, & j=odd,\\
-A_{k^{II}}\delta_{k^{II}}\sin jk^{II}, & j=even.
\end{cases}
\end{eqnarray}
\end{subequations}
respectively. Here the normalization factors are
\begin{equation}
A_{k}=2\left[L+1-\frac{\sin2k\left(L+1\right)}{\sin2k}\right]^{-1/2},
\end{equation}
and \cite{Lieb}
\begin{equation}
\delta_{k}=sgn[\frac{\cos k}{\cos (L_s+1)k}].
\end{equation}
The wave vector $k^I$'s are determined by the equation,
\begin{equation}
\frac{\sin k^{I}\left(L_s+2\right)}{\sin k^{I}L_s}=\frac{U}{t(1+\eta)},
\end{equation}
and $k^{II}$'s are
\begin{equation}
\frac{\sin k^{II}\left(L_s+2\right)}{\sin k^{II}L_s}=\frac{t(1-\eta)}{U}.
\end{equation}

When $|\frac{U}{t(1+\eta)}|>1$, i.e., $\frac{U}{t}> 1+\eta$ or $\frac{U}{t}< -(1+\eta)$, there exists a complex $k_0^{I}$ besides $L_s-1$ real $k$'s (including $\frac{L_s}{2}$ real $k^{II}$'s and $\frac{L_s}{2}-1$ real $k^{I}$'s) \cite{Miao1,Lieb}, which is
\begin{equation}
k_{0}^{I}=\frac{\pi}{2}+iv,
\end{equation}
where $v$ is determined by
\begin{equation}
\frac{\sinh v\left(L_s+2\right)}{\sinh vL_s}=-\frac{U}{t(1+\eta)}.
\end{equation}
For this $k_0^{I}$ mode, the $U$ and $V$ become
\begin{subequations}
\begin{eqnarray}
U_{jk^{I}} & = & \begin{cases}
0, & j = odd,\\
A_{k^{I}}(-1)^{-\frac{j}{2}}\sinh jk^{I}, & j = even,
\end{cases}\\
V_{jk^{I}} & = & \begin{cases}
-A_{k^{I}}(-1)^{\frac{1-j}{2}}\sinh \left(L_s+1-j\right)k^{I}, & j=odd,\\
0, & j = even.
\end{cases}
\end{eqnarray}
\end{subequations}
Then the corresponding normalization factor can be written as,
\begin{equation}
A_{k_{0}^{I}}=2e^{- v L_s}\left(1-e^{-4v}\right)^{1/2},
\end{equation}
and the corresponding singular value is
\begin{equation}
\Lambda_{k_{0}^{I}} \approx (1-|\frac{t(1+\eta)}{U}|)|\frac{t(1+\eta)}{U}|^{L_s/2}.
\label{lambda1}
\end{equation}

In a similar way, when $|\frac{t(1-\eta)}{U}|>1$, i.e., $-(1-\eta) <\frac{U}{t}< 1-\eta$, there exists a complex $k_0^{II}$ that is $k_0^{II}=\frac{\pi}{2}+iv^{'}$ besides $L_s-1$ real $k$'s, where $v^{'}$ is determined by
\begin{equation}
\frac{\sinh v^{'}\left(L_s+2\right)}{\sinh v^{'}L_s}=-\frac{t(1-\eta)}{U}.
\end{equation}
Similarly, we can also obtain the corresponding $U_{jk_{0}^{II}}$, $V_{jk_{0}^{II}}$, $A_{k_0^{II}}$, which equals $A_{k_0^{I}}$, and the corresponding singular value
\begin{equation}
\Lambda_{k_{0}^{II}} \approx (1-|\frac{U}{t(1-\eta)}|)|\frac{U}{t(1-\eta)}|^{L_s/2}.
\label{lambda2}
\end{equation}

%%%%%%%%%%%%%%%%%%%%%%%%%%%%%%%%%%%%%%%
\section{Edge correlation functions and phase diagram}
\label{edge}
\begin{figure}
\includegraphics[height=70mm,width=80mm]{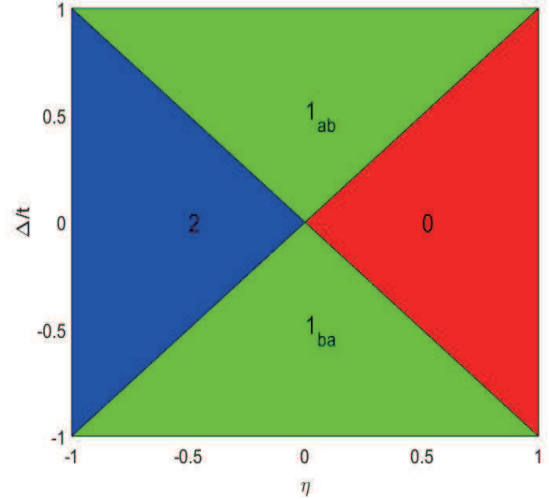}
\caption{\label{01}
  (Color online) The phase labeled by ``0" denotes the SSH-like trivial phase, phases labeled by ``1" express the Kitaev-like topological phases and $2$ denotes the SSH-like topological phase. Here $1_{ab}$ and $1_{ba}$ correspond to different cases with $G_{1L}^{(1)}\neq 0$ and  $G_{1L}^{(2)}\neq 0$  in the thermodynamic limit, respectively.}
\end{figure}

%%%%%%%%%%%%%%%%%%%%%%%%%%%%%%%%%%%%%%
We introduce two edge correlation functions $G_{1L}^{(1)}=\left\langle 0\right|i\gamma_{1}^{a}\gamma_{L_s}^{b}\left|0\right\rangle$ and $G_{1L}^{(2)}=\left\langle 0\right|i\gamma_{1}^{b}\gamma_{L_s}^{a}\left|0\right\rangle$, which can be used to characterize topologically different phases. Before studying the more complicate case of interacting model, we would like to demonstrate that we can reproduce the phase diagram of the non-interacting dimerized Kitaev model by calculating these two edge correlation functions.  We present the phase diagram of this system with $U=0$ and $\mu=0$ in Fig.~\ref{01}, which is consistent with the previous result obtained by calculating the topological numbers of the system with periodical boundary condition \cite{Nagaosa}.  Here the region of "$0$'' in Fig.~\ref{01} corresponds to the case that both $G_{1L}^{(1)}$ and $G_{1L}^{(2)}$ equal zero in the thermodynamic limit, which means that there isn't edge state in this system. Regions labeled by $1_{ab}$ and $1_{ba}$ correspond to the case with $G_{1L}^{(1)}\neq 0$ and $G_{1L}^{(2)}=0$ and the case with $G_{1L}^{(1)}= 0$ and $G_{1L}^{(2)}\neq 0$  in the thermodynamic limit respectively, which means that there exists a Majorana fermion at each end of this chain. Region of "$2$'' corresponds to the case that both $G_{1L}^{(1)}$ and $G_{1L}^{(2)}$ are nonzero in the thermodynamic limit, which means that there exists a Dirac fermion at each end of this chain.

%%%%%%%%%%%%%%%%%%%%%%%%%%%%%%%%%%%%%%%%%%%%
%%%%%%%%%%%%%%%%%%%%%%%%%%%%%%%%%%%%%%%%%%%%%%
We then discuss the interacting dimerized Kitaev model at the symmetric point. Miao et al. \cite{Miao2} have proven that $G_{1L}^{(1)}=\sum_{k}U_{1k}V_{L_sk}$. We then calculate $G_{1L}^{(2)}$,
 \begin{eqnarray}
G_{1L}^{(2)} & = & \left\langle i\gamma_{1}^{b}\gamma_{L_s}^{a} \right\rangle = - \left\langle i \sigma_{1}^{y} \sigma_{L_s}^{x} e^{i\pi \sum_{j=1}^{L_s-1} n_j} \right\rangle  \nonumber \\
 & = & - \left\langle \sigma_{1}^{y} \sigma_{L_s}^{y} e^{i\pi \sum_{j=1}^{L_s} n_j} \right\rangle  \nonumber \\
 & = & -\left\langle \sigma_{1}^{y}\sigma_{L_s}^{y}Z_{2}^{f} \right\rangle = -\left\langle \tilde{\sigma}_{1}^{z}\tilde{\sigma}_{L_s}^{z}Z_{2}^{f} \right\rangle \nonumber \\
 & = & \left\langle \tilde{\gamma}_{1}^{a}\tilde{\gamma}_{1}^{b}\tilde{\gamma}_{L_s}^{a}\tilde{\gamma}_{L_s}^{b}Z_{2}^{f} \right\rangle ,
\end{eqnarray}
where $Z_{2}^{f}$ is the fermion number parity defined as $Z_{2}^{f}=e^{i\pi\sum_l n_l}=(-1)^{N}$ and $N=\sum_l n_l$ is the number of the fermions. It is easy to verify that $(Z_2^{f})^2=1$ and $[H, Z_2^{f}]=0$. For the ground state we can choose $Z_{2}^{f}\left|0\right\rangle =\left|0\right\rangle $ \cite{Miao2}. After using the Wick theorem, we have
 \begin{eqnarray}
G_{1L}^{(2)} & = & \left\langle \tilde{\gamma}_{1}^{a}\tilde{\gamma}_{1}^{b}\right\rangle \left\langle\tilde{\gamma}_{L_s}^{a}\tilde{\gamma}_{L_s}^{b} \right\rangle-\left\langle \tilde{\gamma}_{1}^{a}\tilde{\gamma}_{L_s}^{a}\right\rangle \left\langle\tilde{\gamma}_{1}^{b}\tilde{\gamma}_{L_s}^{b} \right\rangle \nonumber \\
 &&+ \left\langle \tilde{\gamma}_{1}^{a}\tilde{\gamma}_{L_s}^{b}\right\rangle \left\langle\tilde{\gamma}_{1}^{b}\tilde{\gamma}_{L_s}^{a} \right\rangle.
\end{eqnarray}
 where $\left\langle \tilde{\gamma}_{j}^{a}\tilde{\gamma}_{j}^{b}\right\rangle=i\sum_{k}V_{jk}U_{kj}^T=0$ and $\left\langle \tilde{\gamma}_{1}^{a}\tilde{\gamma}_{L_s}^{a}\right\rangle = \sum_{k}U_{1k}U_{kL_s}^T=0$, therefore
\begin{eqnarray}
G_{1L}^{(2)} & = & \left\langle \tilde{\gamma}_{1}^{a}\tilde{\gamma}_{L_s}^{b}\right\rangle \left\langle\tilde{\gamma}_{1}^{b}\tilde{\gamma}_{L_s}^{a} \right\rangle \nonumber \\
 & = & \sum_{k}U_{1k}V_{L_sk}\sum_{k'}V_{1k'}U_{L_sk'}.
\end{eqnarray}
We need investigate the effect of both $k^{I}$ mode and $k^{II}$ mode on the edge correlation functions. Since $U_{1k^{I}}=V_{L_sk^{I}}=0$ and $U_{L_sk^{II}}=V_{1k^{II}}=0$, we have
$G_{1L}^{(1)}=\sum_{k^{II}}U_{1k^{II}}V_{L_sk^{II}}$ and $G_{1L}^{(2)}=\sum_{k^{II}}U_{1k^{II}}V_{L_sk^{II}} \sum_{k^{I}}V_{1k^{I}}U_{L_sk^{I}}$.

When $|U/t|>|1-\eta|$ and $|U/t|<|1+\eta|$, one can easily prove that \cite{Lieb}
\begin{eqnarray}
G_{1L}^{(1)}=\sum_{k^{II}}A_{k^{II}}^{2}\delta_{k^{II}}\sin^{2}k^{II}L_s= O\left(1/L_s\right),
\end{eqnarray}
and $G_{1L}^{(2)}=O\left(1/L_s\right)$, which means that there doesn't exist Majorana fermion at the end of this chain.
When $|U/t|<|1-\eta|$ and $|U/t|<|1+\eta|$, we have
\begin{eqnarray}
G_{1L}^{(1)} & = & \left\langle 0\right| i\gamma_{1}^{a}\gamma_{L_s}^{b} \left|0\right\rangle = U_{1k_{0}^{II}}V_{L_sk_{0}^{II}}+\sum_{k}U_{1k^{II}}V_{L_sk^{II}}\nonumber \\
& = & (-1)^{\frac{L_s}{2}}A_{k_{0}^{II}}^2 \sinh^2v^{'}L_s+\sum_{k^{II}}A_{k^{II}}^2\delta_{k^{II}}\sin^{2}k^{II}L_s \nonumber \\
 & = & (-1)^{\frac{L_s}{2}}\left[1-\left(\frac{U}{t(1-\eta)}\right)^{2}\right]+O\left(1/L_s\right),
\end{eqnarray}
and $G_{1L}^{(2)}=G_{1L}^{(1)}\times O\left(1/L_s\right)$, which equals zero in the thermodynamic limit. Therefore there exists one Majorana fermion at each end of this chain, which corresponds to the TSC phase. When $|U/t|>|1-\eta|$ and $|U/t|>|1+\eta|$, one can easily verify that $G_{1L}^{(1)}=O\left(1/L_s\right)$ and $G_{1L}^{(2)}=G_{1L}^{(1)}\times \{(-1)^{\frac{L_s}{2}}\left[1-\left(\frac{t(1+\eta)}{U}\right)^{2}\right]+O\left(1/L_s\right)\}$, which equals to zero in the thermodynamic limit, which means that there isn't edge state at this chain and the system is trivial.
In a similar way, for this case that $|U/t|>|1+\eta|$ and $|U/t|<|1-\eta|$, we can obtain $G_{1L}^{(1)}\neq 0$ and $G_{1L}^{(2)}\neq 0$ in thermodynamic limit. There exist two Majorana fermions, i.e., one Dirac fermion at each end of this chain, which corresponds to the SSH-like topological phase.
%%%%%%%%%%%%%%%%%%%%%%%%%%%%%%%%%%%%%%%%%%%%
\begin{figure}
\includegraphics[height=65mm,width=80mm]{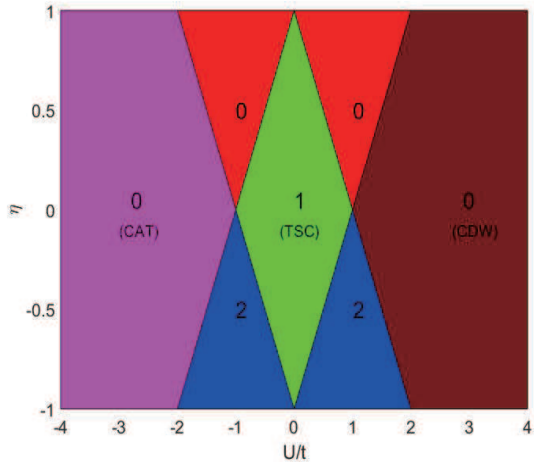}
\caption{\label{02}
  (Color online) Phases labeled by ``0"  denote the trivial phases, where CAT expresses a Shr\"{o}dinger cat-like phase and CDW denotes a charge density wave phase. Phase of ``1"  denotes the topological superconductor (TSC) phase and phases of ``2"  express the SSH-like topological phases.}
\end{figure}
%%%%%%%%%%%%%%%%%%%%%%%%%%%%%%%%%%%%%%%%%%%%%%

%%%%%%%%%%%%%%%%%%%%%%%%%%%%%%%%%%%%%%%%%%%%
\begin{figure}
\includegraphics[height=90mm,width=90mm]{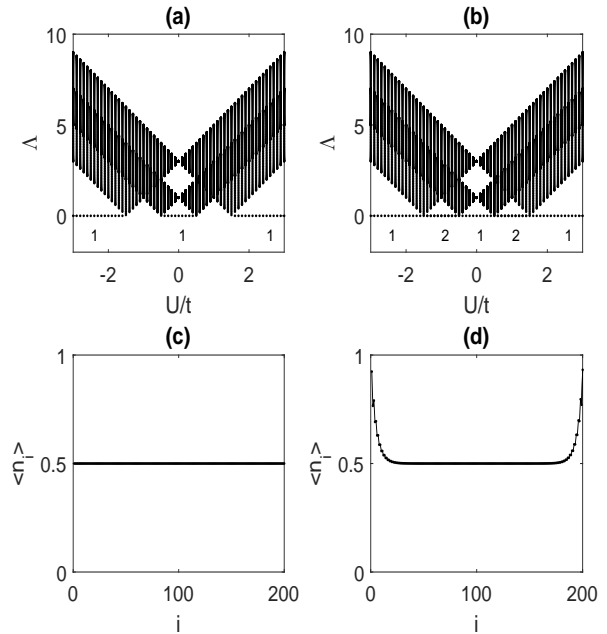}
\caption{\label{03}
   Energy spectra of the interacting dimerized Kitaev chain with (a) $\eta=0.5$, (b) $\eta=-0.5$ and $L_s=200$ as a function of $U/t$ under OBC, where the number of the zero-energy edge states is labeled. Local density distribution of this system with (c) $\eta=0.5, U/t=1$, (d) $\eta=-0.5, U/t=-1$.}
\end{figure}
%%%%%%%%%%%%%%%%%%%%%%%%%%%%%%%%%%%%%%%%%%%%%%

For clear, we display the phase diagram in Fig.~\ref{02}, where regions of ``$0$" denote that there isn't edge state in this chain, which is the trivial phase, region of ``$1$" denotes the TSC phase and there exists a Majorana fermion at each end of this chain and regions of ``$2$'' express the SSH-like topological phase and there exists one Dirac fermion at each end of this chain.
We see that for the $|\eta|< 1$ case, the system is at TSC phase when $|U/t|$ is small, then this system enters into the trivial
phase when increasing the $|U/t|$ to the parameter region $1-\eta <|U/t|< 1+\eta$ for $\eta> 0$ and this system can enter into the SSH-like topological phase if increasing the $|U/t|$ to the region $1+\eta <|U/t|< 1-\eta$ for $\eta<0$. The phase boundary between different phases can be determined by the gap close point in the energy spectrum.

To calculate the energy spectra, we can obtain the singular values $\Lambda$ numerically, which equal to the square of $BB^{T}$ and $B$ is the matrix in Hamiltonian (\ref{ham-3}). Fig.~\ref{03}(a) and (b) show the energy spectra of this system with $\eta=0.5$ and $\eta=-0.5$, respectively. We can also use the density matrix renormalization group (DMRG) method to obtain the ground state $|0 \rangle$ and the local density distribution $\left\langle 0|\hat{n_i}|0 \right\rangle$ of this system in trivial phase and the SSH-like topological phase as shown in Fig.~\ref{03}(c) and (d) respectively. From Fig.~\ref{03}(a), we see that there isn't zero mode at $0.5<|U/t|<1.5$, which is consistent with that it is a trivial insulator phase. The corresponding local density distribution of this phase as shown in fig.~\ref{03}(c), where no edge density distribution is detected in this system. From Fig.~\ref{03}(b), we see that there exist zero modes at $0.5<|U/t|<1.5$ for the $\eta=-0.5$ case, where the zero modes are double degenerate, which can be understood from the Eq.~\ref{lambda1} and Eq.~\ref{lambda2}.
In the thermodynamic limit, $\Lambda_{k_{0}^{I}}$ and $\Lambda_{k_{0}^{II}}$ equal to zero, which means that the $k_{0}^{I}$ mode and $k_{0}^{II}$ mode are the zero modes. Actually, from the above discussion, it is exactly that the $k_{0}^{I}$ mode and the $k_{0}^{II}$ mode give rise to the edge states. Fig.~\ref{03}(d) shows the local density distribution in this SSH-like topological phase and one can see the existence of the edge states.

%%%%%%%%%%%%%%%%%%%%%%%%%%%%%%%%%%%%%%%%%%%%
\begin{figure}
\includegraphics[height=95mm,width=80mm]{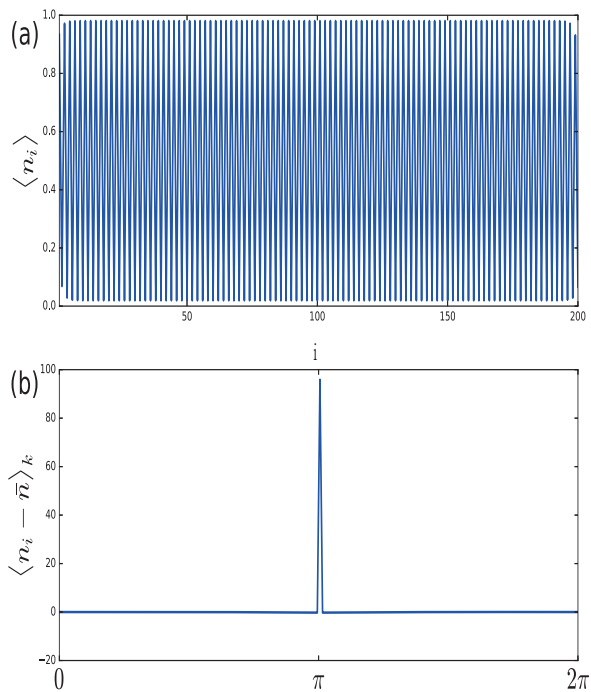}
\caption{\label{04}
   (a) local density distribution and (b) the corresponding Fourier spectrum for this system with $t=1, U=3, \eta=0.5$ and $L_s=200$.}
\end{figure}
%%%%%%%%%%%%%%%%%%%%%%%%%%%%%%%%%%%%%%%%%%%%%%

From Fig.~\ref{03}(a) and (b), we see that there also exist zero energy states for the $|\frac{U}{t}|> |1+\eta|$ and $|\frac{U}{t}| > |1-\eta|$ case.  The zero energy state  corresponds to that $\left\langle i \tilde{\gamma}_{1}^{a}\tilde{\gamma}_{L_s}^{b}\right\rangle=0$ but $\left\langle i \tilde{\gamma}_{1}^{b}\tilde{\gamma}_{L_s}^{a}\right\rangle\neq 0$, which implies the existence of Majorana fermions
$\tilde{\gamma}_{1}^{b}$ and $\tilde{\gamma}_{L_s}^{a}$ at the end of the chain described by the Hamiltonian (\ref{ham-3}). The existence of zero modes at this system holds true even under the unitary transformation of spin rotation. However, for these zero mode states, our results show that $\left\langle i \gamma_{1}^{a}\gamma_{L_s}^{b}\right\rangle=0$ and $\left\langle i \gamma_{1}^{b}\gamma_{L_s}^{a}\right\rangle=0$, which indicates the absence of Majorana edge states in terms of Majorana fermion operators before the spin rotation and these phases should be topologically trivial. To understand these phases, we also show the local density distribution of this system with $U/t=3, \eta=0.5$ and $L_s=200$ in Fig.~\ref{04} (a), which is obtained by using the DMRG method.  From this picture, one can see the local density distributes in an oscillating way corresponding to the charge density wave (CDW) phase \cite{Miao2}.
%and $\left\langle 0|n_{1}|0\right\rangle \approx 1$, $\left\langle 0|n_{L_s}|0\right\rangle \approx 0$, which means that there exists a Dirac fermion at the end of this chain and it is the source of the zero energy modes, but no fermion localize at the other end of this chain, which is consistent with the results obtained by calculating the two edge correlation functions and it is a trivial phase.
Fig.~\ref{04}(b) shows the corresponding Fourier spectrum, which is obtained by taking the fast Fourier transformation of the local density distribution and it is usually used to distinguish the CDW and the incommensurate CDW (ICDW). From this picture, one can see that the Fourier spectrum has a single peak at $\pi$ point and this state is a CDW. For the case that $\frac{U}{t}<-(1+\eta)$, where $\eta > 0$ and $\frac{U}{t}<-(1-\eta)$, where $\eta < 0$, the system is the Shr\"{o}dinger cat-like state with the density distribution being a constant. The Shr\"{o}dinger cat-like state has been studied in Ref. \cite{Miao2}, which is a superposition of two trivial superconductor states with different occupation numbers.

%%%%%%%%%%%%%%%%%%%%%%%%%%%%%%%%%%%%%%%
\section{Summary}
\label{summary}
%%%%%%%%%%%%%%%%%%%%%%%%%%%%%%%%%%%%%%
In summary, we have investigated an exactly solvable interacting dimerized Kitaev model under OBC at the symmetric point of $\Delta=t$ and $\mu=0$ and identified the topological phase diagram by calculating two edge correlation functions and the energy spectra. There exist three different topological phases in various parameter regions: the trivial, TSC and SSH-like topological phases, and the phase boundaries can be determined analytically from the gap close points of the energy spectra. We see that the TSC phase changes to the trivial phase or the SSH-like topological phase when increasing $|U/t|$ and both of them enter into the trivial phase when further increasing the $|U/t|$. For the trivial phase, there also exist three different phases in different parameters regions, i.e., the trivial insulator phase, CDW and CAT phases, which can be distinguished from the energy spectra and the local density distributions. Our results provide a firm ground for further studying and understanding the more general case with $\Delta \neq t$ and  $\mu \neq 0$, for which no exact solution is available but one may calculate the two edge correlation functions numerically.

%%%%%%%%%%%%%%%%
\begin{acknowledgments}
The work is supported by the National Key Research and Development Program of China (2016YFA0300600), NSFC under Grants No. 11425419, No. 11374354 and No. 11174360, and the Strategic Priority Research Program (B) of the Chinese Academy of Sciences  (No. XDB07020000).
\end{acknowledgments}
%%%%%%%%%%%%%%%%%%%%%%%
{\em Note added.} During the preparation of this manuscript, we became aware of a preprint on investigating a similar model \cite{Ezawa} by using a different method. Although the phase diagrams share some similarities, our results show that the CDW phase and the CAT phase are topologically trivial and the zero modes of the two phases aren't Majorana zero modes corresponding to the original fermion operators of this interacting dimerized Kitaev model.
%\appendix
%%%%%%%%%%%%%%%%%%%%%%%%%%%%%e
%\section{}
%%%%%%%%%%%%%%%%%%%%%%%%%%%%%%%%
%In this appendix, we investigate the phase transitions of this system by deriving the topological number.
%\clearpage
%%%%%%%%%%%%%%%%%%%%%%%%%%%%%%%%%%%%%%%%%%%%%

\end{document}